\documentclass[11pt]{article}
\pdfoutput=1
\usepackage[centertags]{amsmath}
\usepackage[square, comma, sort&compress,numbers]{natbib}
\usepackage{array,multirow}
\numberwithin{equation}{section}
\usepackage{amssymb}
\usepackage{graphicx}
\usepackage{color}
\usepackage[normalem]{ulem}

\usepackage{epsfig}
\usepackage{bbold}
\usepackage{pifont}

\usepackage{wrapfig}
\usepackage{float}
\usepackage{soul}

\usepackage{tikz,pgf}
\usetikzlibrary{shapes}
\usetikzlibrary{calc}
\usetikzlibrary{decorations.pathmorphing}
\usetikzlibrary{decorations.pathreplacing,shapes.misc}
\usetikzlibrary{positioning}
\usetikzlibrary{arrows}
\usetikzlibrary{decorations.markings}
\usetikzlibrary{shadings}

\usetikzlibrary{intersections}

\def\sideremark#1{\ifvmode\leavevmode\fi\vadjust{\vbox to0pt{\vss
 \hbox to 0pt{\hskip\hsize\hskip1em
 \vbox{\hsize3cm\tiny\raggedright\pretolerance10000
  \noindent #1\hfill}\hss}\vbox to8pt{\vfil}\vss}}}

\def\be{\begin{equation}}
\def\ee{\end{equation}}
\def\ba{\begin{array}}
\def\ea{\end{array}}

\def\dps{\displaystyle}

\advance\textheight\topskip
\renewcommand{\tilde}{\widetilde}
\renewcommand{\hat}{\widehat}

\newcommand{\bref}[1]{\textbf{\ref{#1}}}

\renewcommand{\geq}{\,{\geqslant}\,}
\renewcommand{\leq}{\,{\leqslant}\,}

\newcommand{\binner}[2]{%
  {\langle}\kern-4.15pt{\langle}#1{,}\,#2{\rangle}\kern-4.15pt{\rangle}}

\newcommand{\half}{\mathchoice{%
    \ffrac{1}{2}}{\frac{1}{2}}{\frac{1}{2}}{\frac{1}{2}}}

\newcommand{\ffrac}[2]{\raisebox{.5pt}%
  {\footnotesize$\displaystyle\frac{#1}{#2}$}\kern1pt}

\def\cC{\mathcal{C}}

\def\cF{\mathcal{F}}
\def\cG{\mathcal{G}}
\def\cH{\mathcal{H}}
\def\cI{\mathcal{I}}

\def\cL{\mathcal{L}}
\def\cM{\mathcal{M}}
\def\cN{\mathcal{N}}
\def\cO{\mathcal{O}}

\def\cS{\mathcal{S}}

\def\cV{\mathcal{V}}

\numberwithin{equation}{section} \makeatletter
\@addtoreset{equation}{section}

\def\be{\begin{equation}}
\def\ee{\end{equation}}
\def\ba{\begin{array}}
\def\ea{\end{array}}

\def\dps{\displaystyle}

\def\1{\tilde{1}}
\def\2{\tilde{2}}
\def\3{\tilde{3}}

\def\tDelta{\tilde\Delta}

\newdimen\tableauside\tableauside=1.0ex
\newdimen\tableaurule\tableaurule=0.4pt
\newdimen\tableaustep
\def\phantomhrule#1{\hbox{\vbox to0pt{\hrule height\tableaurule
width#1\vss}}}
\def\phantomvrule#1{\vbox{\hbox to0pt{\vrule width\tableaurule
height#1\hss}}}
\def\sqr{\vbox{%
  \phantomhrule\tableaustep

\hbox{\phantomvrule\tableaustep\kern\tableaustep\phantomvrule\tableaustep}%
  \hbox{\vbox{\phantomhrule\tableauside}\kern-\tableaurule}}}
\def\squares#1{\hbox{\count0=#1\noindent\loop\sqr
  \advance\count0 by-1 \ifnum\count0>0\repeat}}
\def\tableau#1{\vcenter{\offinterlineskip
  \tableaustep=\tableauside\advance\tableaustep by-\tableaurule
  \kern\normallineskip\hbox
    {\kern\normallineskip\vbox
      {\gettableau#1 0 }%
     \kern\normallineskip\kern\tableaurule}%
  \kern\normallineskip\kern\tableaurule}}
\def\gettableau#1 {\ifnum#1=0\let\next=\null\else
  \squares{#1}\let\next=\gettableau\fi\next}

\def\cC{\mathcal{C}}

\def\cF{\mathcal{F}}
\def\cG{\mathcal{G}}
\def\cH{\mathcal{H}}
\def\cI{\mathcal{I}}

\def\cL{\mathcal{L}}
\def\cM{\mathcal{M}}
\def\cN{\mathcal{N}}
\def\cO{\mathcal{O}}

\def\cS{\mathcal{S}}

\def\cV{\mathcal{V}}

\numberwithin{equation}{section} \makeatletter
\@addtoreset{equation}{section}

\def\be{\begin{equation}}
\def\ee{\end{equation}}
\def\ba{\begin{array}}
\def\ea{\end{array}}

\def\dps{\displaystyle}

\def\ba{\begin{array}}
\def\ea{\end{array}}

\def\dps{\displaystyle}

\usepackage{jheppub}

\makeatletter
\def\@fpheader{\vspace{-.1cm}}
\makeatother

\title{Large-$c$ superconformal torus blocks}

\author[a,b]{Konstantin\ Alkalaev}
\author[a,c,d\,\dagger]{\;Vladimir\ Belavin}
\note[$\dagger$]{Weston Visiting Professorship at Weizmann Institute.}

\affiliation[a]{I.E. Tamm Department of Theoretical Physics, \\P.N. Lebedev Physical
Institute,\\ Leninsky ave. 53, 119991 Moscow, Russia}
\affiliation[b]{Department of General and Applied Physics, \\
Moscow Institute of Physics and Technology, \\
7 Institutskiy per., Dolgoprudnyi, 141700 Moscow region, Russia}
\affiliation[c]{Department of Quantum Physics, \\
Institute for Information Transmission Problems, \\
Bolshoy Karetny per. 19, 127994 Moscow, Russia}
\affiliation[d]{Department of Particle Physics and Astrophysics, Weizmann Institute of Science,\\
Rehovot 7610001, Israel}
\emailAdd{alkalaev@lpi.ru}
\emailAdd{belavin@lpi.ru}

\abstract{We study large-$c$ SCFT$_2$ on a torus  specializing to one-point superblocks in the $\cN=1$ Neveu-Schwarz sector. Considering different contractions of the Neveu-Schwarz  superalgebra related to the large central charge limit we explicitly calculate three superblocks, $osp(1|2)$ global, light, and heavy-light superblocks, and show that they are related to each other. We formulate the $osp(1|2)$ superCasimir eigenvalue equations and identify their particular solutions as the global superblocks. It is shown that the resulting differential equations are the Heun equations. We study exponentiated global superblocks arising at large conformal dimensions and demonstrate  that in the leading approximation the $osp(1|2)$ superblocks are equal to the non-supersymmetric  $sl(2)$  block. 
}
\preprint{FIAN-TD-2018-11}
\arxivnumber{}

\begin{document}

\maketitle
\flushbottom

\section{Introduction}

The study of the large central charge regime in CFT$_2$ allowed to identify conformal blocks as lengths of geodesic networks in the dual gravity theory with matter in asymptotically AdS$_3$ space  \cite{Hartman:2013mia,Fitzpatrick:2014vua,Caputa:2014eta,deBoer:2014sna,Hijano:2015rla,Fitzpatrick:2015zha,Alkalaev:2015wia,Hijano:2015qja,Alkalaev:2015lca} (for further development see \cite{Beccaria:2015shq,Fitzpatrick:2015dlt,Banerjee:2016qca,Alkalaev:2016ptm,Chen:2016dfb,Alkalaev:2016rjl,Kraus:2016nwo,Hulik:2016ifr,Fitzpatrick:2016mtp,Kraus:2017ezw,Belavin:2017atm,Alkalaev:2017bzx,Maxfield:2017rkn,Kusuki:2018wcv,Brehm:2018ipf}).

The aim of this paper is to study  SCFT$_2$ on a torus in the large-$c$ regime.\footnote{We note that for $c\geq 9$ the consistent $\cN=1$ CFT model that comes through the bootstrap condition, is the super-Liouville theory \cite{Belavin:2007gz}.} As a first step in this direction we consider the $\cN=1$ superconformal Neveu-Schwarz (NS) algebra and $1$-point superconformal blocks. We focus on global superblocks that are associated to the $osp(1|2)$ subalgebra of the NS superalgebra. We develop the superCasimir approach where $osp(1|2)$ global superblocks are realized as eigenfunctions of the super-Casimir differential operators. 

The idea is that global blocks in CFT$_2$ on any topology play the central role in studying the large-$c$ regime because all other semiclassical blocks including light and heavy-light blocks  are related to the global one \cite{Fitzpatrick:2015zha,Alkalaev:2015fbw,Alkalaev:2016fok,Cho:2017oxl,Alkalaev:2017bzx}. We show that similar relations are valid in the supercase. Also, we consider the regime of large conformal dimensions and study the exponentiated global superblocks. It is interesting that the supersymmetry turns out to be degenerate in the leading order because the NS supermultiplet conformal dimensions are indistinguishable in this regime, $\Delta +\half\approx \Delta$.  

The paper is organized as follows. Using $osp(1|2)$ representation theory described in Section \bref{sec:rep} we formulate two global superconformal torus blocks corresponding to  different structure constants in Section \bref{sec:block}. The superCasimir approach for torus superblocks is elaborated in Section \bref{sec:casimir}. Here we derive two second order differential equations for the superblocks and analyze their solutions. These equations are in fact the Heun equations and we describe their local and global properties. In Section \bref{sec:exp} the global superblocks are studied in the regime of large conformal dimensions. We show that the resulting superblock block functions  are exponentiated and find explicit expressions. Remarkably, they all are expressed in terms of the exponentiated non-supersymmetric  block function. In Section \bref{sec:contr} we consider superconformal blocks of the NS superalgebra and show that global, light, and heavy-light superblocks can be obtained via particular contractions of the NS superalgebra when $1/c\to 0$.  We close with some concluding remarks in Section \bref{sec:conclusion}. 

\section{Representation theory of $osp(1|2)$ superalgebra}
\label{sec:rep}

In this section we shortly review the $osp(1|2)$ superalgebra and  Verma supermodules. It basically serves to set our notation and conventions. For detailed reviews see, e.g., \cite{Ennes:1997vt,Gotz:2005jz}. 

The superalgebra $osp(1|2)$ is spanned by three even generators $L_{\pm1,0}$ and two odd generators $G_{\pm\half}$  with the graded commutation relations 
\be
\label{osp}
[L_m, L_n] = (m-n)L_{m+n}\;,
\qquad
[L_n,G_{\pm\half}] = \left(\frac{n}{2}\mp\half\right)G_{n\pm\half}\;,
\qquad
\{G_r, G_s\} = 2 L_{r+s}\;, 
\ee
where $m,n = 0,\pm 1$, $r,s = \pm\half$. Obviously, $sl(2)\subset osp(1|2)$.  

\paragraph{Verma supermodule.} Let $\cV_{\Delta}$ denote $osp(1|2)$ supermodule of highest weight $\Delta$. A highest weight (primary) state is defined as 
\be
L_0 |\Delta\rangle  = \Delta |\Delta\rangle\;, \qquad L_1 |\Delta\rangle = 0\;,
\qquad G_{\half} |\Delta\rangle = 0\;.
\ee 
Integer powers of the other basis elements $L_{-1}$ and $G_{-\half}$ act as raising operators generating in this way the supermodule $\cV_{\Delta}$. From  the graded commutation relations \eqref{osp} it follows that $(G_{-\half})^2 = L_{-1}$ and, thus, the supermodule is spanned by 
\be
\label{M}
|M,\Delta\rangle = (L_{-1})^m (G_{-\half})^k |\Delta\rangle\;,
\qquad
M=(m,k)\;:\quad m \in \mathbb{N}_0\;, \quad k=0,1\;.  
\ee 
A number $M = m+k/2$ defines a level, while $k$ defines $\mathbb{Z}_2$ grading of the supermodule. The standard conjugation rules $L_{-m} = (L_m)^\dagger$ and $G_{-s} = (G_s)^\dagger$ are assumed. The $osp(1|2)$ supermodule $\cV_{\Delta}$ has a supermultiplet structure  
\be
\label{supermodule}
\cV_{\Delta} = V_{\Delta}\oplus V_{\Delta+\half}\;,
\ee 
where the factors are $sl(2)$ Verma modules of weights $\Delta$ and $\Delta+\half$. These are  generated from primary states $|\Delta\rangle$ and $|\Delta+\half\rangle$ which are related by a supersymmetry transformation as $|\Delta+\half\rangle = G_{-\half}|\Delta\rangle$.

\paragraph{Invariant operators.} The $osp(1|2)$ superCasimir operator reads 
\be
\label{casimir}
\cS_2 = - L_0^2 + \frac{1}{2}(L_{-1} L_{1}+L_{1} L_{-1})+\frac{1}{4}(G_{-\half}G_{\half}-G_{\half}G_{-\half})\;,
\ee 
with eigenvalues $\cS_2 F = -\Delta(\Delta-\half)F$, where $F$ is a supermultiplet \eqref{supermodule}. It is remarkable that there is the so-called Scasimir operator (see, e.g., \cite{Lesniewski,Arnaudon:1996qe,Ghosh:2003hy})
\be
\label{scasimir}
\Upsilon_2 = \frac{1}{4}(G_{-\half}G_{\half}-G_{\half}G_{-\half}) - \frac{1}{8}\;,
\ee   
that has the property of graded invariance: $[\Upsilon_2, L_{0,\pm1}] = 0$ and $\{\Upsilon_2, G_{\pm \half}\} = 0$ and  satisfies  the quadratic relation $\Upsilon_2 \Upsilon_2  = \cS_2+\frac{1}{64}$ \eqref{casimir}. Then, the Scasimir eigenvalue equation says that 
$\Upsilon_2 F = -\half(\Delta+\frac{1}{4}) F$. Introducing the $sl(2)$ Casimir $C_2$ we can represent the $osp(1|2)$ superCasimir as 
\be
\label{sCd}
\cS_2 = C_2 + \Upsilon_2 +\frac{1}{8}\;.
\ee

\paragraph{Superconformal fields.} Using the superconformal version of the operator-state correspondence, the supermodule \eqref{supermodule} can be realized as a supermultiplet of two conformal operators $(\phi_\Delta(z), \psi_{\Delta+\half}(z))$ with conformal dimensions $\Delta$ and $\Delta+\half$.  Their $osp(1|2)$ superconformal transformations are given by \cite{Friedan:1984rv,Bershadsky:1985dq} 
\be
\label{primaryE}
\begin{aligned}
&[L_m, \phi_\Delta(z)] = z^m(z\partial_z +(m+1)\Delta)\phi_{\Delta}(z)\;,\\
&[G_r, \phi_\Delta(z)] = z^{r+\half} \psi_{\Delta+\half}(z) \;,\\
&[L_m, \psi_{\Delta+\half}(z)] = z^m(z\partial_z +(m+1)(\Delta+\half))\psi_{\Delta+\half}(z)\;,\\
&\{G_r, \psi_{\Delta+\half}(z)\} = z^{r-\half}(z\partial_z +(2r+1)\Delta)\phi_{\Delta}(z)\;,
\end{aligned}
\ee
where $m,n = 0,\pm 1$, $r,s = \pm\half$. It follows that the supermultiplet components are related by a supersymmetry transformation $\psi_{\Delta+\half}(z) = [G_{-\half}, \phi_\Delta(z)]$.

\paragraph{Superfield description.} We use the superfield formalism (see, e.g., \cite{AlvarezGaume:1991bj} and references therein), where $y=(z,\theta)$ are  even and odd holomorphic coordinates in $2|2$ dimensional superplane. Let us consider the (holomorphic) superfield
\be
\Phi_{\Delta}(y) = \phi_{\Delta}(z)+\theta\,\psi_{\Delta+\half}(z)\;,
\quad \text{where}\quad
\psi_{\Delta+\half}(z) = [G_{-\half},\phi_{\Delta}(z)]\;.
\ee
It is assumed that there is a common $\mathbb{Z}_2$ grading $\pi = 0,1$ for coordinates, fields, supermodule  \eqref{supermodule} and superalgebra elements: 
\be
\begin{aligned}
&\pi(\phi_{\Delta}) = 0\;, \quad \pi(\psi_{\Delta+\half})=1\;;
\qquad
\pi(z) = 0\;, \quad \pi(\theta)=1\;;\\
&\pi(V_{\Delta})=0\;,\quad \pi(V_{\Delta+\half}) = 1\;;
\qquad
\pi(L_n)=0\;,\quad \pi(G_{r}) = 1\;.
\end{aligned}
\ee 
Thus, the superfield is even, $\pi(\Phi_{\Delta}(y))=0$.

\section{Global torus superblocks}
\label{sec:block}

Let us consider $osp(1|2)$ superconformal theory on a two-dimensional torus. A one-point function of the primary superfield ${\Phi}_{\Delta, \bar \Delta}(x, \bar  x)$ with conformal dimensions $\Delta, \bar \Delta$ is given by 
\be
\label{1ptf}
\langle {\Phi}_{\Delta, \bar \Delta}(x, \bar  x)  \rangle = str\left[ q^{L_0} \bar{q}^{\bar L_0}{\Phi}_{\Delta, \bar \Delta}(x,\bar x)\right]\;, 
\ee   
where $str$ is the supertrace on the (super)space of states, and $(x,\bar x)=(w,\bar w, \eta, \bar \eta)$ are the supercylindrical coordinates, see Appendix \bref{sec:super}. The modular parameter $q = e^{2\pi i \tau}$, where $\tau\in \mathbb{C}$ is the torus modulus.

\begin{figure}[H]
\centering

\begin{tikzpicture}[line width=1pt]


\draw[blue] (-14,0) circle (1.2cm);
\draw[blue,dashed] (-14,0) circle (1.15cm);

\draw[blue] (-14,-1.2) -- (-14,-3.2);
\fill[blue] (-14,-3.2) circle (0.8mm);

\draw (-13.,-2.8) node {$\phi_\Delta$};


\draw [blue] (-9,0) circle (1.2cm);
\draw [blue,dashed] (-9,0) circle (1.15cm);

\draw[blue,dashed] (-9,-1.2) -- (-9,-3.2);
\fill[blue] (-9,-3.2) circle (0.8mm);

\draw (-11.4,0) node {$(\tDelta,\tDelta+\half) $};
\draw (-8.0,-2.8) node {$\psi_{\Delta+\half}$};

\end{tikzpicture}
\caption{One-point superconformal blocks of the supermultiplet $(\phi_\Delta,\psi_{\Delta+\frac{1}{2}})$. The solid (dashed) lines stand for Grassmann even (odd) operators. The left and right tadpole graphs correspond to the lower block  \eqref{evenblock} and upper block  \eqref{oddblock} and solid-dashed loop represents taking the supertrace.} 
\label{duality}
\end{figure}
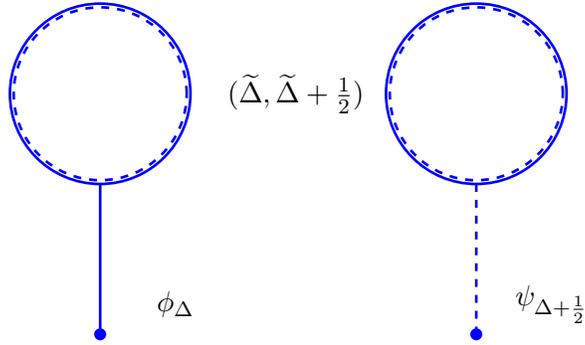

From now on we consider holomorphic sector only. Assuming that the space of states can be decomposed into supermodules of various dimensions $\tilde\Delta$ we can project onto a particular supermodule.  It is convenient to introduce the supertrace function  
\be
\label{def}
B(\Delta, \tilde\Delta, q| x) = str_{_{\tilde\Delta}}\left[ q^{L_0}\Phi_\Delta(x)\right]\;,
\ee
with the supertrace evaluated on the Verma supermodule of weight $\tilde\Delta$.
Then, {\it lower} and {\it upper} superconformal blocks $B_{0}$ and $B_1$ can be defined by means of the decomposition\footnote{Using slightly different conventions superconformal one-point torus blocks can be defined as e.g. in~\cite{Hadasz:2012im}. Related considerations of bosonic one-point blocks can be found in \cite{Poghossian:2009mk,Hadasz:2009db,Piatek:2013ifa,Menotti:2018jsy}.}  
\be
\label{twoblock}
B(\Delta, \tilde\Delta, q| x) = C_{\tDelta \Delta \tDelta}\; B_0(\Delta, \tilde\Delta, q|w)+  C_{\tDelta \Delta+\half \tDelta}\;\eta\, B_1(\Delta, \tilde\Delta, q| w)\;,
\ee   
where  $C_{\tilde \Delta \Delta \tilde \Delta} = \langle \tilde \Delta| \phi_{\Delta}(0) |\tilde \Delta\rangle$ and $C_{\tilde \Delta \Delta+\half \tilde \Delta} = \langle \tilde \Delta| \psi_{\Delta+\half}(0) |\tilde \Delta\rangle$ are two independent structure constant, see relations \eqref{mat} below. Note that  $B_{0,1}$ are even functions.

The lower and upper torus superblocks read off from relations \eqref{def} and \eqref{twoblock} are given by   
\be
\ba{l}
\label{evenblock}
\dps
B_0(\Delta, \tilde\Delta|q) =  \frac{q^{\tDelta}}{\langle \tilde \Delta| \phi_{\Delta}(w) |\tilde \Delta\rangle}\Bigg[\sum_{m=0}^{\infty} \, q^m\,
\frac{\langle \tilde \Delta, m| \phi_{\Delta}(w) |m, \tilde \Delta\rangle}{\langle \tilde \Delta, m |m, \tilde \Delta\rangle}-
\\
\\
\dps
\hspace{3cm}\;\;-\sum_{m=0}^{\infty} \, q^{m+\half}\,
\frac{\langle \tilde \Delta+\half, m| \phi_{\Delta}(w) |m, \tilde \Delta+\half\rangle}{\langle \tilde \Delta+\half, m |m, \tilde \Delta+\half\rangle}\Bigg]\;,
\ea
\ee
\be
\ba{l}
\label{oddblock}
\dps
B_1(\Delta, \tilde\Delta|q) =  \frac{q^{\tDelta}}{\langle \tilde \Delta| \psi_{\Delta+\half}(w) |\tilde \Delta\rangle}\Bigg[\sum_{m=0}^{\infty} \, q^m\,
\frac{\langle \tilde \Delta, m| \psi_{\Delta+\half}(w) |m, \tilde \Delta\rangle}{\langle \tilde \Delta, m |m, \tilde \Delta\rangle}+
\\
\\
\dps
\hspace{3cm}\;\;+\sum_{m=0}^{\infty} \, q^{m+\half}\,
\frac{\langle \tilde \Delta+\half, m| \psi_{\Delta+\half}(w) |m, \tilde \Delta+\half\rangle}{\langle \tilde \Delta+\half, m |m, \tilde \Delta+\half\rangle}\Bigg]\;.
\ea
\ee
The superblocks \eqref{evenblock} and \eqref{oddblock} are decomposed according to parity of the exchanged channel that is manifested by respectively first and second sums of each expression (see Fig. \bref{duality}). Taking into account the form of the matrix elements in the even/odd sectors, 
\be
\label{mat}
\begin{aligned}
&\langle \tilde \Delta+\half|\tilde \Delta+\half\rangle = 2\tilde \Delta \langle \tilde \Delta|\tilde \Delta\rangle = 2\tilde \Delta \;,\\
&\langle \tilde \Delta+\half| \phi_{\Delta}(z) |\tilde \Delta+\half\rangle = (2\tilde\Delta - \Delta)\langle \tilde \Delta| \phi_{\Delta}(z) |\tilde \Delta\rangle\;,\\
&\langle \tilde \Delta| \phi_{\Delta}(z) |\tilde \Delta+\half\rangle = -\langle \tilde \Delta| \psi_{\Delta+\half}(z) |\tilde \Delta\rangle\;,\\
&\langle \tilde \Delta+\half| \psi_{\Delta+\half}(z) |\tilde \Delta+\half\rangle = -(2\tilde \Delta + \Delta-\half)\langle \tilde \Delta| \psi_{\Delta+\half}(z) |\tilde \Delta\rangle\;,
\end{aligned}
\ee
we find the closed expressions for the lower superblock function,  
\be
\label{superblock0}
\ba{l}
\dps
q^{-\tDelta} B_0(\Delta, \tilde\Delta|q) =  \frac{1}{(1-q)^{\Delta}} \;\,{}_2 F_{1}(2\tilde \Delta - \Delta, 1-\Delta, 2\tilde \Delta\, |\, q)-
\\
\\
\dps 
\hspace{3cm}- \frac{2\tilde\Delta - \Delta}{2\tilde\Delta}\frac{q^{1/2}}{(1-q)^{\Delta}} \;\,{}_2 F_{1}(2\tilde \Delta - \Delta+1, 1-\Delta, 2\tilde \Delta+1\, |\, q)\;,
\ea
\ee
and for the upper superblock function,
\be
\label{superblock1}
\ba{l}
\dps
q^{-\tDelta}B_1(\Delta, \tilde\Delta|q) = \frac{1}{(1-q)^{\Delta+\half}} \;\,{}_2 F_{1}(2\tilde \Delta - \Delta-\half, -\Delta+\half, 2\tilde \Delta\, |\, q)-
\\
\\
\dps 
\hspace{2cm}- \frac{2\tilde \Delta + \Delta-\half}{2\tilde\Delta}\frac{q^{1/2}}{(1-q)^{\Delta+\half}} \;\,{}_2 F_{1}(2\tilde \Delta - \Delta+\half, -\Delta+\half, 2\tilde \Delta+1\, |\, q)\;.
\ea
\ee
 
Sending $\Delta\to 0$ in \eqref{superblock0} we find the $osp(1|2)$ (graded) character 
\be
\label{character}
q^{-\tDelta}B_0(0, \tilde\Delta|q) = \frac{1}{1+q^{1/2}} = 1-q^{1/2}+q-q^{3/2}+q^2-q^{5/2}+q^3+ \ldots\;.
\ee
Indeed, substituting $\Phi_{\Delta} = \mathbb{1}$ into \eqref{def} yields the supertrace of the identity operator.  In accordance with \eqref{M} it shows that there is one state on each level of the supermodule $\cV_{\tilde\Delta}$ \eqref{supermodule}. Note that $B_0(0, \tilde\Delta|q) = B_1(1/2, \tilde\Delta|q)$ and $B_0(1/2, \tilde\Delta|q) = B_1(0, \tilde\Delta|q)$. 

Equivalently, the $osp(1|2)$ character can be obtained by expanding both superblock functions near  $\tDelta = \infty$,  
\be
\label{tDelta}
B_0(\Delta, \infty|q) = B_1(\Delta, \infty|q) = B_0(0, \tilde\Delta|q) \;.
\ee

\section{SuperCasimir eigenvalue equations }
\label{sec:casimir}

It is known that $CFT_2$ global blocks can be described as solutions to the second-order differential equations interpreted as the $sl(2)$ Casimir operator eigenvalue conditions imposed on exchanged channels \cite{Dolan:2011dv}. The original construction for 4-point blocks on the sphere can be extended to higher-point conformal blocks on the sphere and torus \cite{Alkalaev:2015fbw,Kraus:2017ezw}. The superCasimir equations for $4$-point sphere blocks were previously discussed in \cite{Fitzpatrick:2014oza,Bobev:2015jxa,Cornagliotto:2017dup}.  In $d$ dimensions (super)conformal Casimir equations were discussed in \cite{SimmonsDuffin:2012uy,Fitzpatrick:2014oza,Bobev:2015jxa,Cornagliotto:2017dup,Gobeil:2018fzy,Poland:2018epd}.\footnote{We note that $o(d+1,1)$ algebra in Euclidean CFT$_d$ describes global conformal symmetries. In CFT$_2$ on general Riemannian  surfaces Virasoro algebra always contains $o(3,1)\approx sl(2,\mathbb{C})$ subalgebra that describes global symmetry only on the sphere.} 

In what follows we elaborate the superCasimir approach for torus superblocks. To this end, we note that acting with the superCasimir operator \eqref{casimir} inside the supertrace operation we get the eigenvalue equation for the exchanged channel  
\be
\label{1eigen}
str_{_{\tilde\Delta}}\left[\cS_2\, q^{L_0}\Phi_\Delta(x)\right] = -\tilde\Delta(\tilde\Delta-\half)B(\Delta, \tilde\Delta, q| x)\;,
\ee
where the factor on the right-hand side is the eigenvalue of the superCasimir on the irreducible Verma supermodule $\cV_{\tDelta}$. On the other hand, the external superfield $\Phi_{\Delta}$ satisfies the other eigenvalue equation    
\be
\label{2eigen}
\tilde\cS_2 \Phi_{\Delta}(x) = -\Delta(\Delta-\half) \Phi_{\Delta}(x)\;,
\ee
with the superCasimir operator given by  
\be
\label{diffS}
\tilde \cS_2 = - \cL_0^2 + \frac{1}{2}(\cL_{-1} \cL_{1}+\cL_{1} \cL_{-1})+\frac{1}{4}(\cG_{-\half}\cG_{\half}-\cG_{\half}\cG_{-\half})\;,
\ee
where the $osp(1|2)$ generators are realized as differential operators in the supercylindrical coordinates, see Appendix \bref{sec:super}. Using the operator-state correspondence we can show that the eigenvalue superCasimir equation \eqref{2eigen} is identically satisfied. Combining two eigenvalue conditions \eqref{1eigen} and \eqref{2eigen} we will  obtain two equations for two superblocks. We will see that the supertrace function embracing two superblocks \eqref{twoblock} provides a natural way to impose the superconformal invariance conditions. 

\subsection{Derivation of the eigenvalue equation}

Using the approach of \cite{Kraus:2017ezw} we find the following identities with the $osp(1|2)$ basis elements inserted into the supertrace 
\be
\label{iden}
\ba{l}
\dps
str_{_{\tilde\Delta}}\left[L_{-n}L_n\, q^{L_0}\Phi_\Delta(x)\right] = \left(\frac{2n q^n}{1-q^n} q\frac{\partial}{\partial q}+ \frac{1}{(1-q^n)(1-q^{-n})} \cL_{-n}\cL_{n}\right)str_{_{\tilde\Delta}}\left[q^{L_0}\Phi_\Delta(x)\right]\;,
\\
\\
\dps
str_{_{\tilde\Delta}}\left[G_{-k}G_k\, q^{L_0}\Phi_\Delta(x)\right] = \left(-\frac{2q^k}{1-q^k} q\frac{\partial}{\partial q}+ \frac{1}{(1-q^k)(1-q^{-k})} \cG_{-k}\cG_{k}\right)str_{_{\tilde\Delta}}\left[q^{L_0}\Phi_\Delta(x)\right]\;,
\ea
\ee
where $n=\pm1$ and $k = \pm \half$. Here, in particular, we used the commutator $[L_0, T_s] = -s T_s$ with  $T_{s} = (L_{0,\pm1}, G_{\pm\half})$ being   $osp(1|2)$ basis elements that yields the identity $T_s q^{L_0} = q^{L_0+s} T_s$. Also, it is easy to derive the polynomial homogeneity identity 
\be
\label{L02}
str_{_{\tilde\Delta}}\left[(L_0)^m q^{L_0}\Phi_\Delta(x)\right] = \left(q\frac{\partial}{\partial q}\right)^m str_{_{\tilde\Delta}}\left[q^{L_0}\Phi_\Delta(x)\right]\;,\qquad m=0,1,2,...\;.
\ee
Loosely speaking, the identities \eqref{iden} and \eqref{L02} convert polynomial combinations of basis elements $L_n$ and $G_k$ acting on states of $\cV_{\tDelta}$ to differential operators $\cL_n$ and  $\cG_k$ acting on the superfield along with $q$-differential operators. It is crucial here that the  superblocks are combined into the supertrace function that allows using the graded cyclic property and commuting even/odd operators via (graded) cyclic permutations.  

Recalling  the $U(1)\times U(1)$ global symmetry of two-dimensional torus and relation \eqref{comcyl} we find that $\cL_0$ acts trivially, i.e., 
\be
\label{L01}
\cL_0 \,str_{_{\tilde\Delta}}\left[q^{L_0}\Phi_\Delta(x)\right] = 0\;\;.
\ee
Then, substituting the identities \eqref{iden} and  \eqref{L02} along with the relation \eqref{L01} into the superCasimir  equation \eqref{1eigen} we get 
\be
\label{interCas}
\begin{aligned}
&\left[(q\partial_q)^2 +\left(\frac{1}{2} - \frac{1}{1+q^{1/2}}\right)(q\partial_q)+
\frac{q}{(1-q)^2} \tilde C_2 + \frac{q^{1/2}}{(1-q^{1/2})^2}\left(\tilde\Upsilon_2+\frac{1}{8}\right)-\right.\\
&\dps\hspace{8cm}-\left.\tilde\Delta\left(\tilde\Delta-\frac{1}{2}\right)\right]str_{_{\tilde\Delta}}\left[ q^{L_0}\Phi_\Delta(x)\right] = 0\;,
\end{aligned}
\ee
where operators $\tilde C_2$ and $\tilde\Upsilon_2$ are directly read off from \eqref{sCd} and \eqref{diffS}. 
Now, we have to know how $\tilde C_2$ and $\tilde\Upsilon_2$ act on the superfield $\Phi_{\Delta}(x)$. To this end, we use the superCasimir equation \eqref{2eigen} to express  
\be
\left(\tilde\Upsilon_2 +\frac{1}{8}\right)\Phi_{\Delta} =-\Delta(\Delta-\half) \Phi_{\Delta} - \tilde C_2 \Phi_{\Delta}\;,
\ee
and substitute this relation into the equation \eqref{interCas}
\be
\label{interCas2}
\begin{aligned}
&\left[(q\partial_q)^2 +\left(\frac{1}{2} - \frac{1}{1+q^{1/2}}\right)(q\partial_q)+
\left(\frac{q}{(1-q)^2} -\frac{q^{1/2}}{(1-q^{1/2})^2}\right)\tilde C_2 -\Delta(\Delta-\half) \frac{q^{1/2}}{(1-q^{1/2})^2}-\right.\\
&\dps\hspace{8cm}-\left.\tilde\Delta\left(\tilde\Delta-\frac{1}{2}\right)\right]str_{_{\tilde\Delta}}\left[ q^{L_0}\Phi_\Delta(x)\right] = 0\;.
\end{aligned}
\ee
Now, we observe that the $sl(2)$ Casimir operator acts non-diagonally on the supermodule
\be
\tilde C_2 \Phi_{\Delta}(x) = -\Delta(\Delta-1)\phi_\Delta(w) -(\Delta+\half)(\Delta-\half)\,\eta\, \psi_{\Delta}(w)\;.
\ee
To derive this relation we used formulas from the Appendix \bref{sec:super} and the fact that even/odd components of the $osp(1|2)$ superfield are themselves  $sl(2)$ conformal fields. Substituting this relation into \eqref{interCas2} and using  \eqref{twoblock} we finally find two equations
\be
\label{Fineq1}
\ba{l}
\dps
\left[(q\partial_q)^2 +\left(\frac{1}{2} - \frac{1}{1+q^{1/2}}\right)(q\partial_q)-\Delta(\Delta-1)
\left(\frac{q}{(1-q)^2} -\frac{q^{1/2}}{(1-q^{1/2})^2}\right) -\right.
\\
\\
\dps
\hspace{55mm}\left.-\Delta(\Delta-\half) \frac{q^{1/2}}{(1-q^{1/2})^2}-\tilde\Delta(\tilde\Delta-\half)\right]B_0(\Delta, \tilde\Delta| q) = 0\;,
\ea
\ee
\be
\label{Fineq2}
\ba{l}
\dps
\left[(q\partial_q)^2 +\left(\frac{1}{2} - \frac{1}{1+q^{1/2}}\right)(q\partial_q)-(\Delta+\half)(\Delta-\half)
\left(\frac{q}{(1-q)^2} -\frac{q^{1/2}}{(1-q^{1/2})^2}\right) -\right.
\\
\\
\dps
\hspace{55mm}\left.-\Delta(\Delta-\half) \frac{q^{1/2}}{(1-q^{1/2})^2}-\tilde\Delta(\tilde\Delta-\half)\right]B_1(\Delta, \tilde\Delta| q) = 0\;.
\ea
\ee
These are the second order ODEs that differ only in the coefficient of the third terms. Since each of the equations has two independent solutions we fix the asymptotics  
\be
B_{0,1}(\Delta, \tilde\Delta| q) \to q^{\tDelta}\qquad \text{as}\qquad q\to 0\;, 
\ee
and find that the corresponding solutions to the superCasimir equations \eqref{Fineq1} and \eqref{Fineq2} are given by functions \eqref{superblock0} and \eqref{superblock1}.

\subsection{Properties of the eigenvalue equation}

One might wonder whether the local OPE data and the modular properties of the torus correlation functions fix the form of the differential equations \eqref{Fineq1} and \eqref{Fineq2}. To clarify this issue,  we change  variables as $x=q^{1/2}$, for example, in the first equation \eqref{Fineq1} to obtain the second order  ODE
\be 
\label{fuchsian}
B_0''(x)+\frac{2}{x+1}B_0'(x)-\frac{2 \left(\Delta x \left(2 \Delta x+x^2+1\right)+2 \tDelta^2 \left(x^2-1\right)^2-\tDelta \left(x^2-1\right)^2\right)}{x^2 \left(x^2-1\right)^2}B_0(x)=0\;,
\ee
(the prime denotes $x$-derivative) with the (regular) singular points $-1,0,1$ and $\infty$, and the Riemann P-symbol 
\begin{equation}
\label{Riem}
\begin{Bmatrix}  
 x=-1 & & x=0& & x=1 & & x=\infty  \\ 
\Delta-1 && 1-2 \tDelta && \Delta+1 && 1-2 \tDelta  \\
-\Delta && 2 \tDelta && -\Delta && 2 \tDelta  \\
\end{Bmatrix}.
\end{equation}
From the entities $\alpha_{i,j}$  of the P-symbol (which are two characteristic exponents, $j=1,2$ at singular points $x_i$, where index $i=-1,0,1,\infty$ labels singular points) 
one can  recognize  the superCasimir  eigenvalues in the exchanged channel \eqref{1eigen}. 
The equation \eqref{fuchsian} is Fuchsian that can be seen, in particular, by checking the  Fuchs identity $\sum_{i,j}\alpha_{i,j} =(p-2) n(n-1)/2$, where $p(=4)$ is the number of singular points 
and $n(=2)$ is the order of the ODE.

An important characteristic of the Fuchsian ODEs is the so-called rigidity index 
\be
\cI= n^2 (2-p)+ \sum_{i,j}  m_{i,j}^2\;,
\ee
where $m_{i,j}$ are multiplicities of the characteristic exponents $\alpha_{i,j}$ \eqref{Riem}. In our case the spectral type $(\{m_{i,j}\})=(1,1;1,1;1,1;1,1)$. 
If the rigidity index $\cI =2$, then the local data completely defines  a differential equation, in particular, its explicit form and the monodromy group, etc. (for details see, e.g., \cite{oshima}).   
In our case, however, the rigidity index $\cI=0$ implying that the ODE is not rigid and contains one accessory parameter which  form cannot be determined  only from the local data \eqref{Riem}.

Doing the transformation  $B_0(x) =   (1 - x)^{\alpha_{1,2}}(1 + x)^{\alpha_{-1,1}}x^{\alpha_{0,2}} y(x)$ we get the differential equation of the form
\be\label{Heuneq}
y''(x)+\left(\frac{a+b-c-d+1}{x-t}+\frac{c}{x}+\frac{d}{x-1}\right)y'(x) +\frac{(a b x-s)}{x (x-1) (x-t)}y(x) =0\;.
\ee
This is the Heun equation with singular point $0,1,t,\infty$, and the  parameters $a,b,c,d$, along with  the accessory parameter $s$ expressed in terms of conformal dimensions. The solution to the equation \eqref{Heuneq} is the Heun function $\text{Hn}({t,s};{a,b,c,d};x)$, see Appendix \bref{sec:heun}.\footnote{In CFT$_2$ with Virasoro symmetry the Heun equation arises as the $c \to \infty$ limit of the BPZ equation for the 5-point conformal blocks with one degenerate light operator \cite{Litvinov:2013sxa} (see also recent discussion in \cite{Piatek:2017fyn,Lencses:2017dgf}).} 
In our case, $t=-1$ and  $s=2 \Delta (4 \tDelta-1)$. Matching other parameters one arrives at the differential equation of the form 
\be
y''(x)+\frac{4 (\tDelta x^2-\Delta x-\tDelta)}{x^3-x}y'(x)+\frac{\Delta (2-8 \tDelta) }{x^3 -x}y(x)=0\;.
\ee 
Hence, the lower superblock can be expressed in terms of the Heun function 
\be
\label{Hn0}
\begin{aligned}
&B_0(\Delta, \tDelta| q) =(1-q^{1/2})^{-\Delta} (1+q^{1/2})^{\Delta-1} q^{\tDelta}\times\\
&\hspace{4cm}\times \text{Hn}(-1,2 \Delta (4 \tDelta-1)\,;\, 0,4 \tDelta-1,4 \tDelta,-2 \Delta\,;\,q^{1/2})\;.
\end{aligned}
\ee 

The superCasimir equation for the upper superblock \eqref{Fineq2} can be considered along the same lines. The resulting Heun's representation is given by 
\be
\label{Hn1}
\begin{aligned}
&B_1(\Delta, \tDelta| q) =\left(1-q^{1/2}\right)^{\Delta-\frac{1}{2}} \left(1+q^{1/2}\right)^{-\Delta-\frac{1}{2}} q^{\tDelta}\times\\
&\hspace{4cm}\times \text{Hn}(-1,(1-2 \Delta) (4 \tDelta-1)\,;\, 0,4 \tDelta-1,4 \tDelta,2 \Delta-1\,;\,q^{1/2})\;.
\end{aligned}
\ee

\subsection{Exponentiated  global superblocks}
\label{sec:exp}

Let us consider large conformal dimensions. In this regime  $\Delta$ and $\tilde \Delta$ tend to infinity uniformly 
\be
\label{kappa}
\Delta = \kappa\,  \sigma\;, \qquad \tilde\Delta  = \kappa \,\tilde \sigma\;,
\ee
where the scale parameter $\kappa \gg 1$, and  $\sigma$ and $\tilde\sigma$ are {\it finite} rescaled conformal dimensions. Remarkably, the large rescaled  dimensions $\Delta+\half$ and $\tilde\Delta+\half$ are also given by $\sigma$ and $\tilde \sigma$.

Given that  $\sigma$ and $\tilde \sigma$ are fixed,  the large-$\kappa$ asymptotics of the superblocks \eqref{superblock0} and \eqref{superblock1} can be conveniently defined using the superCasimir equations \eqref{Fineq1}, \eqref{Fineq2}. Indeed, in this case the solutions are  exponentiated,
\be
\label{expo}
q^{-\tDelta}\,B_\alpha(\Delta, \tilde\Delta, q) \sim \exp\left[\sum_{n=0}^\infty \,\kappa^{- n+1}\, b_{\alpha|n}(\sigma, \tilde \sigma,q)\right]\;, \qquad \alpha = 0,1\;,
\ee 
where $b_{\alpha|0}(\sigma, \tilde\sigma, q)$ are the {\it classical} global blocks, while $b_{\alpha|n}(\sigma, \tilde \sigma,q)$ at $n = 1,2,...$ are $1/\kappa^n$ corrections. Below we show that the leading asymptotics coincide $b_{0|0}(\sigma, \tilde\sigma, q) = b_{1|0}(\sigma, \tilde\sigma, q)$ and the difference arises in higher orders. We note that exponential factors can be represented as follows   
\be
\label{delta}
b_{\alpha|n}(\sigma, \tilde \sigma,q) = \tilde\sigma^{1-n}\, b_{\alpha|n}(\delta,q)\;,
\qquad \delta = \frac{\sigma}{\tilde \sigma} \;,
\ee
where the lightness parameter $\delta$ is dimensionless. This formula means that the scale factor $\kappa$ in \eqref{expo} is dummy because $\kappa^{- n+1}\, b_{\alpha|n}(\sigma, \tilde \sigma,q) = \tDelta^{-n+1}b_{\alpha|n}(\Delta/\tDelta,q)$. However, working with dimensionless parameters $\kappa$ and $\delta$ turns out to be  technically more convenient.  

Substituting the ansatz \eqref{expo} into the superCasimir equations we find out that each leading asymptotic satisfies the same residual differential equation 
\be
\label{quadra}
\left(\partial_q b_{\alpha|0}(\sigma, \tilde\sigma, q)\right)^2+\frac{2\tilde\sigma}{q}\,\partial_q b_{\alpha|0}(\sigma, \tilde\sigma, q)-\frac{\sigma^2}{q(1-q)^2}=0\;.
\ee
Therefore, classical lower/upper superblocks coincide with each other and equal to the known  classical $sl(2)$ global block \cite{Alkalaev:2016fok}
\be
\label{leading}
b_{\alpha|0}(\sigma, \tilde\sigma, q) = \tilde\sigma \int_0^q dx\left( -\frac{1}{x}+\sqrt{\frac{1}{x^2}+\frac{\delta^2}{x(x-1)^2}}\,\right)\;,
\qquad
\alpha = 1,2\;.
\ee
The integral is divergent in $x=0$. However, the divergence can be bypassed if we expand in the lightness parameter  \eqref{delta} near $\delta=0$ meaning that the exchanged channel is much heavier than  external primary operator (see \cite{Alkalaev:2016fok} for more details). 

All higher-order corrections are iteratively expressed in terms of the leading contribution \eqref{leading}. For example, the $\cO(k^{0})$ corrections $b_{\alpha|1}(\sigma, \tilde\sigma, q)$ are defined by inhomogeneous  first order differential equations given in Appendix \bref{sec:corrections}. The solutions have the integral form
\be
\label{sol1}
\ba{c}
\dps
b_{0|1}(\sigma, \tilde\sigma, q) =\int_0^q dx\left[\frac{(x+1) (x-1)^2-2 W x^{1/2}}{4 W (1-x) x}+\frac{(x-1)^2-  (x+1) x^{1/2}\delta}{4 W^{1/2} (x-1) x}\right]\;,
\ea
\ee  
\be
\label{sol2}
\ba{c}
\dps
\;\;\;b_{1|1}(\sigma, \tilde\sigma, q) =\int_0^q dx\left[\frac{(x+1) (x-1)^2-2 W x^{1/2}}{4 W (1-x) x}+\frac{\left(x^{1/2}+1\right)^2 \left(x+1+(\delta -2) x^{1/2}\right)}{4 W^{1/2} (x-1) x}\right]\;,
\ea
\ee  
where $W = (1-x)^2 +x \delta^2$. The above integrals differ in the last two terms. Evaluating the integrals and further expanding in the smallness parameter $\delta$ we can obtain a first few terms
\be
\label{decompose}
\begin{aligned}
&b_{0|1}(\sigma, \tilde\sigma, q) = -\log \left(1+q^{1/2}\right)+\frac{1}{2}\frac{q^{1/2}}{(1-q)}\delta-\frac{1}{8}\frac{q (q+1)}{ (1-q)^2}\delta^2+\cO(\delta^3)\;,\\
&b_{1|1}(\sigma, \tilde\sigma, q) =-\log \left(1+q^{1/2}\right)-\frac{1}{2}\frac{q^{1/2}}{ \left(1-q^{1/2}\right)}\delta-\frac{1}{8}\frac{q (q+1)}{ (1-q)^2}\delta^2
+\cO(\delta^3)\;.
\end{aligned}
\ee
In particular, recalling that these corrections are dimensionless \eqref{delta} we observe that $osp(1|2)$ character \eqref{character} is reproduced by the logarithmic  terms  in the limit $\delta \to 0$. Indeed, the limit can be achieved by $\tilde\sigma \to \infty$ and therefore we can use \eqref{tDelta}.   

\section{Conformal blocks of contracted Neveu-Schwarz  superalgebras}
\label{sec:contr}

In this section we consider the Inonu-Wigner contractions of the NS superalgebra with respect to the inverse central charge $1/c$ in the limit $c\to \infty$.   Analogously  to the Virasoro algebra case \cite{Alkalaev:2016fok} we compute associated  torus superblocks and identify them with different types of semiclassical torus superblocks.\footnote{On the sphere, $\cN=1,2$  vacuum 4-point  heavy-light superblocks were considered in \cite{Chen:2016cms}; 4-point light superblocks were calculated  in \cite{Poghosyan:2017qdv}; the global $\cN=1,2$ superblocks were discussed in \cite{Fitzpatrick:2014oza}; extended SCFT superblocks, including $\cN=4$ case can be found in \cite{Lin:2015wcg}.  The torus NS superblocks were studied in \cite{Hadasz:2012im}.}

\subsection{NS superalgebra and superblocks}

Let us consider $\cN=1$ NS superalgebra which generators $L_m$ and $G_r$ satisfy the graded commutation relations 
\be
\ba{l}
\dps[L_m, L_n] = (m-n)L_{m+n} + \frac{c}{12} (m^3-m)\delta_{m+n,0}\;,
\\
\dps
[L_n,G_r] = \left(\frac{n}{2}-r\right)G_{n+r}\;,
\\
\dps
\{G_r, G_s\} = 2 L_{r+s} +\frac{c}{3} \left(r^2-\frac{1}{4}\right)\delta_{r+s,0}\;,

\ea
\ee
where $m,n \in \mathbb{Z}$ and $r,s \in \mathbb{Z}+1/2$. A primary superfield $\Phi_{\Delta}(x)$ transformations on a torus  are given in  \eqref{comcyl}. 

We introduce the supertrace function on the NS supermodule of the weight $\tDelta$ and define (cf. Section \bref{sec:block}) the NS superblocks as 
\be
\label{defNS}
str_{_{\tilde\Delta}}\left[ q^{L_0-\frac{c}{24}}\Phi_\Delta(x)\right] \equiv \Upsilon(\Delta, \tilde\Delta, q| x) = C_{\tDelta \Delta \tDelta}\, \Upsilon_0(\Delta, \tilde\Delta, q|w)+  C_{\tDelta \Delta+\half \tDelta}\,\eta \Upsilon_1(\Delta, \tilde\Delta, q| w)\;,
\ee
where $C_{\tDelta \Delta \tDelta}$ and $C_{\tDelta \Delta+\half \tDelta}$ are the structure constants, and the component form of the NS lower/upper superblocks reads\footnote{For further convenience, we omitted the overall prefactor $q^{-c/24}$. See \cite{Hadasz:2012im} for extended discussion of NS superblocks.}
\be
\label{Upsilon}
\Upsilon_0(\Delta,\tilde\Delta,c | q) =\sum_{n\in\half \mathbb{Z}}(-)^{2n}\,q^{\tDelta+n}\sum_{n = |\mathsf{M}|=|\mathsf{N}|} B^{\mathsf{M}|\mathsf{N}}\,
\frac{\langle \tilde \Delta, \mathsf{M}| \phi_{\Delta}(w) |\mathsf{N}, \tilde \Delta\rangle}{\langle \tilde \Delta| \phi_{\Delta}(w) | \tilde \Delta\rangle }\;,
\ee 
\be
\label{Upsilon1}
\Upsilon_1(\Delta,\tilde\Delta,c | q) =\sum_{n\in\half \mathbb{Z}}\,q^{\tDelta+n}\sum_{n = |\mathsf{M}|=|\mathsf{N}|} B^{\mathsf{M}|\mathsf{N}}\,
\frac{\langle \tilde \Delta, \mathsf{M}| \psi_{\Delta+\half}(w) |\mathsf{N}, \tilde \Delta\rangle}{\langle \tilde \Delta| \psi_{\Delta+\half}(w) |\tilde \Delta\rangle}\;.
\ee 
Here, summation over superindices is not graded and basis monomials of the NS supermodule denoted by $\cM_{\tDelta}$ are given by 
\be
\label{moduleM}
|\mathsf{N},\tDelta\rangle =  L_{-m_1}^{i_1} \cdots L_{-m_k}^{i_k}   G_{-s_1}^{j_1} \cdots L_{-s_l}^{j_l}  |\tDelta\rangle\;, 
\ee
where $\mathsf{N} = (K,S)$ labels basis monomials, $|\mathsf{N}| = |K|+|S|$, where $|K| = i_1m_1+ \ldots +i_k m_k$ and $|S| = j_1s_1+ \ldots +j_l s_l$ is the level number. The supermatrix $B^{\mathsf{M}|\mathsf{N}}$ is the inverse of the Gram supermatrix
$B_{\mathsf{M}|\mathsf{N}} = {\langle \tilde \Delta, \mathsf{M} |\mathsf{N}, \tilde \Delta\rangle}$. 

In what follows we consider various semiclassical superblocks arising in the limit $c\to\infty$ of the original NS superblocks \eqref{Upsilon}, \eqref{Upsilon1}. We distinguish between heavy and light conformal dimensions $\Delta = \cO(c^1)$ and $\Delta = \cO(c^0)$ so that there are  three types of one-point superblocks with external light operators: global superblock, light superblock, and heavy-light superblock. It will be shown that these superblocks are associated to different contractions of the NS superalgebra and are related to each other. 

\subsection{Contracted NS superalgebras}

Let us first consider contractions that leave a finite-dimensional subalgebra intact. Even and odd generators rescale as 
\be
\begin{aligned}
&L_{0, \pm 1} \to l_{0,\pm 1} =  L_{0, \pm 1} \;, \qquad\quad L_{m} \to a_m =  L_{m}/c^\gamma\;, \quad  |m|\geq 2\;,\\
&\;G_{\pm \half} \to g_{\pm \half} =  G_{\pm \half} \;, \qquad\quad\quad G_{r} \to b_r =  G_{r}/c^\gamma\;, \quad \; |r|\geq 3/2\;.
\end{aligned}
\ee
We consider two cases:  type A contraction $\gamma = 1$ and type B  contraction $\gamma = 1/2$.   The rescaled transformations of the primary field $\Phi_\Delta(x)$ with respect to $l_n$ and $g_s$ take the form \eqref{primaryE}, while higher order NS generators act trivially,
\be
\label{singlet}
[a_m, \Phi_{\Delta}] = 0\;, 
\qquad
[b_s, \Phi_{\Delta}] = 0\;.
\ee
Thus, $\Phi_\Delta$ is an $osp(1|2)$ superconformal quasi-primary field.

In general, the resulting contracted NS superalgebras are isomorphic to semi-direct sum  
\be
\label{contra}
osp(1|2) \ltimes \cF\;,
\ee
where $\cF$ is an infinite-dimensional superalgebra with two branches $\cF = \cF_- \oplus \cF_+$ spanned by  basis elements $a_{m}$ and $b_{n}$ with $n\in \half \mathbb{Z}$. The two branches $\cF_{\pm}$ are highest weight $osp(1|2)$ supermodules. The factor $\cF$ is defined by a particular type A/B contraction.

\paragraph{Type A contraction.} In this case the contracted superalgebra is
\be
\label{typeA}
NS_{_{A}} = osp(1|2)\,\ltimes \,\cal{SA}\;,
\ee
where $\cal{SA}$ is an infinite-dimensional Abelian superalgebra. The   graded commutation relations are given by 
\be
\label{typeA1}
[l_{m}, l_n]  = (m-n)l_{m+n}\;, 
\qquad
\{g_r, g_s\} = 2 l_{r+s}\;,
\qquad
[l_n, g_r] =(\frac{n}{2}-r) g_{n+r}\;,
\ee
\be
\label{typeA2}
[a_m, a_n] = 0\;,\qquad \{b_r, b_s\} = 0\;,
\qquad
[a_n, b_r] = 0\;,
\ee 
\be
\label{typeA3}
\begin{aligned}
&[l_m, a_n] = (m-n)a_{m+n}\;, \qquad |m+n|\geq 2\;;
\qquad
[l_m, a_n] = 0\;, \qquad |m+n|\leq 1\;,\\
&[l_n, b_r] = (\frac{n}{2}-r) b_{n+r}\;,\qquad |n+r| \geq \frac{3}{2}\;;
\qquad
[l_n, b_r] = 0\;,\qquad |n+r|=\half\;,\\
&[g_r,a_n] = -(\frac{n}{2}-r) b_{n+r}\;,
\qquad
\{g_r, b_s\}  = 2 a_{r+s}\;,\quad |r+s|\geq 2\;;
\qquad 
\{g_r, b_s\}  = 0 \;, \quad |r+s|\leq 1\;.
\end{aligned}
\ee
Here, the first group of relations defines $osp(1|2)$ superalgebra, the second group defines  Abelian superalgebra, the third group defines the highest weight supermodule structure. 

\paragraph{Type B contraction.} Here, the contracted superalgebra is 

\be
\label{typeB}
NS_{_{B}} = osp(1|2)\,\ltimes \,\cal{HC}\;,
\ee
where $\cal{HC}$ is the infinite-dimensional Heisenberg-Clifford superalgebra. The respective  graded commutation relations are different from those of $NS_A$ \eqref{typeA1}--\eqref{typeA3} only in the part defining the infinite-dimensional factor. Namely, 
\be
\label{typeB1}
\ba{c}
\dps
[a_m, a_n] = \frac{m(m^2-1)}{12}\delta_{m+n,0}\;,
\qquad
\{b_r,b_s\} = \frac{1}{3} \left(r^2-\frac{1}{4}\right)\delta_{r+s,0}\;,
\qquad
[a_n, b_r] = 0\;.
\ea
\ee

\paragraph{Type C contraction.} Also, there is a third type of contraction when all NS generators  are rescaled so that the $osp(1|2)$ subalgebra is also contracted,
\be
L_{0} \to l_0 =  L_{0}/c\;,
\qquad
L_{m} \to l_m =  L_{m}/\sqrt{2c}\;, 
\qquad 
G_{r} \to g_r = G_{r}/\sqrt{2c}\;, 
\qquad 
m\neq 0\;,\;\; \forall r\;.
\ee 
Then, the resulting contraction is given by the Heisenberg-Clifford superalgebra 
\be
\label{typeC}
NS_{_{C}} = \cal{HC}\;,
\ee
with non-zero  graded commutation relations
\be
\label{HCrels}
\begin{aligned}
&[l_m, l_{-m}] = m \,l_0+\frac{m(m^2-1)}{24}\;,\qquad m \in \mathbb{Z}\;,\\
&\{g_r, g_{-r}\} = l_0+\frac{1}{6} \left(r^2-\frac{1}{4}\right)\;,\qquad r \in \mathbb{Z}+\half\;.
\end{aligned}
\ee
The superfield $\Phi_\Delta$ is light and, therefore, $\cH\cC$-invariant,
\be
[l_m, \Phi_\Delta] = 0\;,
\qquad
[g_{r},\Phi_\Delta] = 0\;,
\qquad
m\in \mathbb{Z}\;,\quad r \in \mathbb{Z}+\half\;.
\ee

\subsection{Associated superblocks} 

In what follows we argue that the superblock functions associated to the three types of contracted NS superalgebra correspond to $osp(1|2)$ global, light, and heavy-light one-point superblocks, respectively. 

Consider first the representation theory of the contracted superalgebras \eqref{contra}. A supermodule  $\cM_{\tDelta}$ is spanned by basis monomials  
\be
\label{conbasis}
|\mathsf{N}, \tilde \Delta\rangle  =  a_{\bar R} b_{\bar S}\, l_{-1}^s \,g_{-\half}^k\,|\tilde \Delta\rangle \equiv a_{\bar R} b_{\bar S}\, |N,\tilde \Delta\rangle \;, 
\ee
where we split into  $osp(1|2)$ and $\cF$ subalgebra generators, and  denoted 
$a_{\bar R} = a_{-m_1}^{i_1} \ldots  a_{-m_k}^{i_k}$, with a level $|\bar R| = i_1m_1 + \dots +i_k m_k$, and $b_{\bar S} = b_{-s_1}^{j_1} \ldots  b_{-s_k}^{j_k}$, with a level $|\bar S| = j_1s_1 + \dots +j_k s_k$. The total level number is given by $|\mathsf{N}| = |N|+ |\bar R|+|\bar S| \in \half\mathbb{Z}$, where $N$ labels $osp(1|2)$ indices and $|N| = s+k$ and  $s =1,2,...$ and $k=0,1$, cf. \eqref{M} and \eqref{moduleM}. 

Let  $\cV_{\tDelta}$ be an $osp(1|2)$ supermodule of weight $\tDelta$, and $F$ be the  Fock supermodule of the factor $\cF$ in the truncated superalgebra  \eqref{contra}. Then, considering $\cM_{\tDelta}$, $\cV_{\tDelta}$, and $F$ on their own as linear spaces we conclude from \eqref{conbasis} that $\cM_{\Delta}$ is the (graded) tensor product of vector spaces  $\cM_{\tDelta} =F \otimes \cV_{\tDelta}$.
On the other hand, the primary superfield $\Phi_\Delta$ is $\cF$-invariant \eqref{singlet} 
and, therefore, can be represented as $\mathbb{1}_{_{F}}\otimes \hat\Phi_\Delta$, where $\hat \Phi_\Delta$ is $osp(1|2)$ quasi-primary superfield. Using that a supertrace trace on the tensor product is a product of supertraces one can explicitly show that the supertrace function \eqref{defNS} associated to the contracted type A/B superalgebras reads  
\be
\label{traces}
\Upsilon(\Delta,\tilde\Delta,q|\eta) = \chi_{_F}(q) \cdot str_{_{\cV_{\tDelta}}}\left[ q^{L_0}\hat\Phi_\Delta(x)\right]\;,
\ee
where the first factor here is the $\cF$ character on the Fock module, while the second factor is the $osp(1|2)$ supertrace function \eqref{twoblock}. 

For the type A superalgebra \eqref{typeA} the $\cF$ character is trivial $\chi_{_F}(q)=1$ so that  the resulting supertrace function is $\Upsilon(\Delta,\tilde\Delta,q|\eta) = B(\Delta,\tilde\Delta,q|\eta)$ given in  \eqref{def}. Thus, we conclude that global superblocks correspond to the type A contracted NS superalgebra. In other words, a truncation of the NS superalgebra to  $osp(1|2)$ subalgebra is equivalent to a particular contraction.  

For the type B superalgebra \eqref{typeB} the $\cF$ character is a truncated Heisenberg-Clifford (graded) character. Indeed, the Heisenberg-Clifford character is known to be (see, e.g., \cite{Goddard:1986ee}) 
\be
\label{HC}
\chi_{_{\cal{HC}}}(q) = \prod_{n=0}^\infty \frac{1+q^{n+\half}}{1-q^{n+1}}\;.
\ee
On the other hand, basis elements $a_m$ and $b_r$ of the $\cF$ factor are labeled by (half-)integers $|m| \geq  2$ and $|r|\geq 3/2$. It means that states generated by basis elements with $m=0,\pm 1$ and $r = \pm 1/2$ do not contribute because they belong to the $osp(1|2)$ module. The $\cF$ character, if compared to the Heisenberg-Clifford character, does not take into account those lower label states so that the truncated character $\tilde{\chi}_{_{\cal{HC}}}(q)$ is given by 
\be
\label{trunc}
\tilde{\chi}_{_{\cal{HC}}}(q) = \frac{\chi_{_{\cal{HC}}}(q)}{\chi_{_{osp}}(q)}  = \prod_{n=1}^\infty \frac{1-q^{n+\half}}{1-q^{n+1}}\;,
\ee
where the $osp(1|2)$ character given by \eqref{character}. Finally, the resulting block is 
\be
\label{light}
\cL(\Delta,\tilde\Delta,q|\eta) = \tilde{\chi}_{_{\cal{HC}}}(q) B(\Delta, \tDelta,q|\eta)\;,
\ee
where the prefactor is given by \eqref{trunc}. The block on the left-hand side is the light NS superblock  that equivalently can be obtained as the $c\to \infty $ limit of the original NS superconformal block \eqref{defNS} at fixed conformal dimensions $\Delta, \tDelta$. 

Finally, let us shortly consider heavy-light superblocks $\cH_{0,1}(\Delta,\tilde\Delta,q)$ which have one heavy dimension $\tDelta = \cO(c^1)$ and light dimension $\Delta = \cO(c^0)$ at $c\to \infty$. A  contraction of the NS superalgebra underlying the heavy-light superblocks is given by the type C superalgebra \eqref{typeC}. Applying the definitions \eqref{Upsilon} and \eqref{Upsilon1} to the  type C  we can explicitly calculate the associated superblocks that are given by the Heisenberg-Clifford character \eqref{HC}, i.e.  
\be
\label{HLblock}
\cH_{0,1}(\Delta,\tilde\Delta,q) = \chi_{_{\cal{HC}}}(q)\;.
\ee
Also, using relations \eqref{tDelta}, \eqref{trunc}, and \eqref{light} we see that the heavy-light superblocks can be seen as the limiting case of the light  superblocks
\be
\label{HLvsL}
\cH(\Delta,\tilde\Delta,q|\eta) = \cL(\Delta,\infty,q|\eta)\;.
\ee 

The above relations between various superblocks \eqref{light}, \eqref{HLblock}, and \eqref{HLvsL} are a supersymmetric version of the analogous relations between semiclassical Virasoro torus block and $sl(2)$ global torus blocks \cite{Alkalaev:2016fok}.\footnote{For analogous relations between higher-point global and light non-supersymmetric torus blocks see also \cite{Cho:2017oxl}. On the sphere, it is shown that 4-point heavy-light  $\cN=1$ vacuum superblocks with pairwise equal dimensions coincide with non-supersymmetric heavy-light blocks in the leading order of the large-$c$ approximation  \cite{Chen:2016cms}.} 

\section{Concluding remarks}
\label{sec:conclusion}

In this paper we have developed a framework to study large-$c$ behavior of torus SCFT$_{2}$ superblocks. We have explicitly calculated  various types of semiclassical one-point superblocks and established relations  between them.  
 
We have seen that two exponentiated $osp(1|2)$ global superblocks are equal to the single non-supersymmetric exponentiated  $sl(2)$ global block. On the other hand, there is a lot of evidence that exponentiated global blocks in the leading approximation are related to the linearized classical conformal blocks \cite{Fitzpatrick:2015zha,Alkalaev:2015fbw,Alkalaev:2016fok,Alkalaev:2017bzx}. We expect that in SCFT$_2$ these two types of superblocks are similarly related and, therefore, the linearized classical superblocks are equal to the linearized non-supersymmetric classical blocks. Indeed, here we have the same phenomenon  that classical conformal dimensions of the supermultiplet operators $\epsilon_b = \frac{\Delta}{c}$ and $\epsilon_f = \frac{\Delta+\half}{c}$ coincide in the large-$c$ regime: $\epsilon_b = \epsilon_f $. It is similar to the argument of \cite{Hadasz:2007nt} that classical $\cN=1$  superblocks on the sphere should have the  large-$c$ asymptotic given by the purely bosonic Zamolodchikov's classical block.  

From the bulk perspective,  both classical global and linearized classical torus blocks are realized as lengths of geodesic tadpole-type  networks stretched in the thermal AdS$_3$ space \cite{Alkalaev:2016ptm,Kraus:2017ezw,Alkalaev:2017bzx}. One might expect that classical superblocks could be realized in terms of superparticles propagating on the particular background that solves $3d$ supergravity equations.\footnote{In that context, it would be instructive to examine geodesic Witten diagrams with fermions \cite{Nishida:2018opl}.} However, presently, we  may conclude only that in the leading order of the large-$c$ approximation the superblocks are realized by duplicated system of bosonic geodesic networks.  

The other interesting question is to understand the bulk dual realization of higher-order corrections to the classical global (super)blocks of Section \bref{sec:exp}. It appears that using the worldline approach they can be calculated by accounting for the backreaction of particles in the bulk (see \cite{Maxfield:2017rkn} for recent discussion of the worldline formalism in the context of the semiclassical AdS$_3$/CFT$_2$ correspondence).   

\vspace{5mm}
\noindent \textbf{Acknowledgements.} We thank R. Metsaev for useful discussions. The work of K.A. was supported by the Russian Science Foundation grant 14-42-00047.

\appendix

\section{Supercylindrical coordinates}
\label{sec:super}
The superfield transformations on the superplane are known to be \cite{Friedan:1984rv,Bershadsky:1985dq} 
\be
\begin{aligned}
&[L_n,\Phi_{\Delta}(y)]  = \left(z^{n+1} \partial_z + \half(n+1)z^n  \theta \partial_\theta + \Delta(n+1)z^n\right) \Phi_{\Delta}(y)\;,\\
&[G_r,\Phi_{\Delta}(y)]  = \left(z^{r+\half} (\partial_\theta - \theta \partial_z) - 2\Delta(r+\half) \theta z^{r-\half}\right) \Phi_{\Delta}(y)\;,  
\end{aligned}
\ee
where $n\in \mathbb{Z}$ and $r\in \mathbb{Z}-\half$.
To find the superconformal transformations on the torus we change from the superplane coordinates $y=(z,\theta)$ to the supercylindrical coordinates 
$x=(w,\eta)$ as
\be
w = i \log z\;, \qquad  \eta  = (-i z)^{-1/2} \theta\;. 
\ee 
This is the superconformal map so that  the superfield $\Phi_\Delta(y)$ transforms as\footnote{For more details on $2d$ superconformal geometry see, e.g., \cite{Friedan,Dorrzapf:1997rx}.} 
\be
\Phi_{\Delta}(y) = (D\eta)^{2\Delta} \Phi_{\Delta}(x)\;,\quad \text{where} 
\quad 
D\eta = (-i z)^{-1/2}\;, 
\ee
and $D = \partial_\theta + z\partial_z$ is the supercovariant (left) derivative.
In the supercylindrical coordinates we find
\be
\label{comcyl}
\begin{aligned}
&\cL_n\Phi_{\Delta}(x) \equiv [L_n,\Phi_{\Delta}(x)]  = (- i)e^{-inw}\left[ \partial_w +i n\left(\half \eta\partial_\eta +\Delta\right)\right] \Phi_{\Delta}(x)\;,\\
&\cG_r \Phi_{\Delta}(x) \equiv  [G_r,\Phi_{\Delta}(x)]  = (-i)^{-\half}e^{-irw}\left[ \left(\partial_\eta + \eta\partial_w \right)+2ir\Delta \eta \right]\Phi_{\Delta}(x)\;.   
\end{aligned}
\ee
As expected, the Hamiltonian on a cylinder is realized by $w$-translations, $\cL_0 = -i \partial_\omega$ contrary to dilatations $\cL_0 = z\partial_z + \Delta$ on the plane.

\section{More on the Heun function }
\label{sec:heun}

A power series solution of the Heun equation \eqref{Heuneq} that goes to $1$ at $x=0$ has the expansion \cite{Erdel} 
\be
\text{Hn}({t,s};{a,b,c,d};x) = \sum_{n=0}^\infty \alpha_n x^n\;,
\ee
where $\alpha_0 = 1$ and other expansion coefficients are defined by the  system of recurrent equations  
\be
t c \alpha_1 - s \alpha_0 = 0 \;,
\qquad
R_n \alpha_{n+1}-(Q_n	+s)\alpha_n + P_n\alpha_{n-1} = 0\;,
\qquad n=1,2,...\;,
\ee
where 
\be
\begin{aligned}
&R_{n} = (n-1+a)(n-1+b)\;,\\
& Q_n = n((n-1+c)(1+t)+td+(a+b-c-d+1))\;,\\
& P_n = t(n+1)(n+c)\;.
\end{aligned}
\ee
\section{Corrections to the classical global block}
\label{sec:corrections}

The first corrections $b_{\alpha|1} = b_{\alpha|1}(\sigma, \tilde\sigma, q)$ to the classical global block are defined by the following differential equations  
\be
\label{eq12}
\partial_q b_{0|1} = \frac{A}{B}\;,
\qquad
\partial_q b_{1|1} = -\frac{C}{D}\;,
\ee
where 
\be
\begin{aligned}
&A = -2 (q-1)^2 q^{3/2} \partial^2_q b_{0|0}-\left(q^{1/2}-1\right)^2 \left(3 q^{3/2}+4 q+q^{1/2}\right) \partial_q b_{0|0}+2 \left(-q^{3/2}+q+q^{1/2}-1\right) \tilde\sigma+(q+1)\sigma\;,\\
&B = 4 (q-1)^2 q^{1/2} \left(q \partial_q b_{0|0}+\tilde\sigma\right)\;,\\
&C = q^{1/2} (q^{1/2}-1)^2 \left(2 \left(q^{1/2}+1\right) q \partial^2_q b_{0|0}+\left(3 q^{1/2}+1\right) \partial_q b_{0|0}\right)+\left(q^{1/2}+1\right) \sigma+2 \left(q^{1/2}-1\right)^2 \tilde\sigma\;,\\
&D = 4 (q^{1/2}-1)^2 (q+q^{1/2}) \left(q \partial_q b_{0|0}+\tilde\sigma\right)\;,
\end{aligned}
\ee
and the classical global block $b_{0|0}$ is given by \eqref{leading}. The solutions are given in \eqref{sol1}-\eqref{decompose}.

Below we give the first correction to the $sl(2)$ global block \eqref{leading}  included here for completeness,  
\be
b_1(\sigma,\tilde\sigma,q) = -\log (1-q) +\frac{1}{2}\frac{q}{(q-1)}\delta-\frac{1}{4}\frac{q^2}{(q-1)^2}\delta^2  + \frac{1}{24}\frac{(q-3) q^2}{(q-1)^3}\delta^3 +\cO(\delta^4)\;.
\ee
It can be read off from the Casimir equation in \cite{Alkalaev:2016fok,Kraus:2017ezw}. Similar to the superblocks \eqref{decompose}, the first term here yields the $sl(2)$ character. 

\providecommand{\href}[2]{#2}\begingroup\raggedright\endgroup

\end{document}